% hep-th/0403050
% last edited by alex, 3/3
% ssb.tex

\input harvmac
\input epsf

%%%%%%%%%%%%%%%%%%%%%%%%%%  FIGURES   %%%%%%%%%%%%%%%%%%%%%%%%%%%%%%%

\newcount\figno
\figno=0
\def\fig#1#2#3{
\par\begingroup\parindent=0pt\leftskip=1cm\rightskip=1cm\parindent=0pt
\baselineskip=11pt \global\advance\figno by 1 \midinsert
\epsfxsize=#3 \centerline{\epsfbox{#2}} \vskip 12pt {\bf Fig.\
\the\figno: } #1\par
\endinsert\endgroup\par
}
\def\triplefig#1#2#3#4#5{
\par\begingroup
\global\advance\figno by 1 \epsfxsize=#5
a) \hskip 2.7in b)
\par{\epsfbox{#2}} \hskip 0.2in {\epsfbox{#3}}
\par
c)
\par {\epsfbox{#4}}
\vskip 12pt {\bf Fig.\ \the\figno: } #1 \endgroup\par
}
\def\figlabel#1{\xdef#1{\the\figno}}
\def\encadremath#1{\vbox{\hrule\hbox{\vrule\kern8pt\vbox{\kern8pt
\hbox{$\displaystyle #1$}\kern8pt} \kern8pt\vrule}\hrule}}
\def\p{\partial}

\def\Re{{\rm Re}\,}
\def\Im{{\rm Im}\,}
\def\t{\tau}

%\SenNU
\lref\SenNU{
A.~Sen,
``Rolling tachyon,''
JHEP {\bf 0204}, 048 (2002)
[arXiv:hep-th/0203211].
%%CITATION = HEP-TH 0203211;%%
}

%\MaloneyCK
\lref\MaloneyCK{
A.~Maloney, A.~Strominger and X.~Yin,
``S-brane thermodynamics,''
JHEP {\bf 0310}, 048 (2003)
[arXiv:hep-th/0302146].
%%CITATION = HEP-TH 0302146;%%
}
%\DowkerGB
\lref\DowkerGB{
F.~Dowker, J.~P.~Gauntlett, G.~W.~Gibbons and G.~T.~Horowitz,
``The Decay of magnetic fields in Kaluza-Klein theory,''
Phys.\ Rev.\ D {\bf 52}, 6929 (1995)
[arXiv:hep-th/9507143].
%%CITATION = HEP-TH 9507143;%%
}

 %\LambertZR
\lref\LambertZR{
N.~Lambert, H.~Liu and J.~Maldacena,
``Closed strings from decaying D-branes,''
arXiv:hep-th/0303139.
%%CITATION = HEP-TH 0303139;%%
}
%\GaiottoRM
\lref\GaiottoRM{
D.~Gaiotto, N.~Itzhaki and L.~Rastelli,
``Closed strings as imaginary D-branes,''
arXiv:hep-th/0304192.
%%CITATION = HEP-TH 0304192;%%
}

\lref\Weylpaper{H.~Weyl, Ann.~Phys.~(Leipzig) {\bf 54} (1917) 117}
%\EmparanWK
\lref\EmparanWK{
R.~Emparan and H.~S.~Reall,
%``Generalized Weyl solutions,''
Phys.\ Rev.\ D {\bf 65}, 084025 (2002)
[arXiv:hep-th/0110258].
%%CITATION = HEP-TH 0110258;%%
}
%\MaldacenaRE
\lref\MaldacenaRE{
J.~M.~Maldacena,
%``The large N limit of superconformal field theories and supergravity,''
Adv.\ Theor.\ Math.\ Phys.\  {\bf 2}, 231 (1998)
[Int.\ J.\ Theor.\ Phys.\  {\bf 38}, 1113 (1999)]
[arXiv:hep-th/9711200].
%%CITATION = HEP-TH 9711200;%%
}

%\HorowitzCD
\lref\HorowitzCD{
G.~T.~Horowitz and A.~Strominger,
%``Black Strings And P-Branes,''
Nucl.\ Phys.\ B {\bf 360}, 197 (1991).
%%CITATION = NUPHA,B360,197;%%
}

\lref\IsraelKhan{W.~Israel and K.~A.~Khan, Nuovo Cim., {\bf 33},
331 (1964).}

%\SenMG
\lref\SenMG{
A.~Sen,
``Non-BPS states and branes in string theory,''
arXiv:hep-th/9904207.
%%CITATION = HEP-TH 9904207;%%
}

%\EmparanAU
\lref\EmparanAU{
R.~Emparan,
``Black diholes,''
Phys.\ Rev.\ D {\bf 61}, 104009 (2000)
[arXiv:hep-th/9906160].
%%CITATION = HEP-TH 9906160;%%
}
%\TeoUG
\lref\TeoUG{
E.~Teo,
``Black diholes in five dimensions,''
Phys.\ Rev.\ D {\bf 68}, 084003 (2003)
[arXiv:hep-th/0307188].
%%CITATION = HEP-TH 0307188;%%
}
%\EmparanBB
\lref\EmparanBB{
R.~Emparan and E.~Teo,
``Macroscopic and microscopic description of black diholes,''
Nucl.\ Phys.\ B {\bf 610}, 190 (2001)
[arXiv:hep-th/0104206].
%%CITATION = HEP-TH 0104206;%%
} \lref\gary{G. Gibbons, private communication.}
%\EmparanGM
\lref\EmparanGM{
R.~Emparan and M.~Gutperle,
``From $p$-branes to fluxbranes and back,''
JHEP {\bf 0112}, 023 (2001)
[arXiv:hep-th/0111177].
%%CITATION = HEP-TH 0111177;%%
}
%\GutperleAI
\lref\GutperleAI{
M.~Gutperle and A.~Strominger,
``Spacelike branes,''
JHEP {\bf 0204}, 018 (2002)
[arXiv:hep-th/0202210].
%%CITATION = HEP-TH 0202210;%%
}
%\Liang
\lref\Liang{
Y.~C.~Liang and E.~Teo,
``Black diholes with unbalanced magnetic charges,''
Phys.~Rev.~D {\bf 64} (2001)
[arXiv:hep-th/0101221].
%%CITATION = HEP-TH 0101221 ;%%
}
%\Peet
\lref\Peet{
A.~W.~Peet,
``TASI Lectures on Black Holes in String Theory,''
[arXiv:hep-th/0008241].
%%CITATION = HEP-TH 0008241 ;%%
}
%\GrossPerry
\lref\GrossPerry{
D.~J.~Gross and M.~J.~Perry,
``Magnetic Monopoles in Kaluza-Klein Theories,''
Nucl.~Phys.~B226 (1983), p.~29.}
%\Sorkin
\lref\Sorkin{
R.~D.~Sorkin,
``Kaluza-Klein Monopole,''
Phys.~Rev.~Let. {\bf 51} 2, p. 87.}
%\GibbonsHawking
\lref\GibbonsHawking{
G.~W.~Gibbons and S.~W.~Hawking,
``Classification of Gravitational Instanton Symmetries,''
Commun.~Math.~Phys.~{\bf 66} (1979) p. 291.}

\lref\herlt{E. Herlt, ``Static and Stationary Axially Symmetric Gravitational
Fields of Bounded Sources. I. Solutions Obtainable from the van Stockum Metric,''
Gen. Rel. Grav., Vol. 9, No. 8, p. 711.}

\lref\herlttwo{E. Herlt, ``Static and Stationary Axially Symmetric Gravitational
Fields of Bounded Sources. II. Solutions Obtainable from Weyl's Class,''
Gen. Rel. Grav., Vol. 11, No. 5, p. 337.}

\lref\sibgatullin{N. R. Sibgatullin, {\it Oscillations and Waves
in Strong Gravitational and Electromagnetic Fields}, Springer-Verlag,
Berlin Heidelberg 1991.  (Originally {\it Kolebaniya i volny v silnykh gravitatsionnykh i elektromagnitnykh polyakh}, Nauka, Moscow 1984.)}

\lref\mankofirst{V. S. Manko, J. Mart\'{\i}n, E. Ruiz, ``Metric of two
arbitrary Kerr-Newman sources located on the symmetry axis,''
J. Math Phys. {\bf 35} (12), December 1994, p. 6644.}

\lref\mankosecond{V. S. Manko, J. Mart\'{\i}n, E. Ruiz, ``Extended
family of the electrovac two-soliton solutions for the Einstein-Maxwell
equations,'' Phys. Rev. D {\bf 51} (8), April 1995, p. 4187.}

\lref\mankothird{V. S. Manko, J. Mart\'{\i}n, E. Ruiz, ``Extended
$N$-soliton solution of the Einstein-Maxwell equations,''
Phys. Rev. D {\bf 51} (8), April 1995, p. 4192.}

\lref\myers{R. C. Myers, ``Higher-dimensional black holes in compactified
space-times,'' Phys. Rev. D {\bf 35} (2), January 1987, p. 455.}

%\lref\emparan{R. Emparan, ``Black diholes,'' {\tt hep-th/9906160}.}

%\lref\emparanteo{R. Emparan and E. Teo, ``Macroscopic and Microscopic
%Description of Black Diholes,'' {\tt hep-th/0104206}.}

\lref\ernstfirst{F. J. Ernst, ``New Formulation of the
Axially Symmetric Gravitational Field Problem,'' Phys. Rev. {\bf 167} (5),
25 March 1968, p. 1175.}

\lref\ernstsecond{F. J. Ernst, New Formulation of the
Axially Symmetric Gravitational Field Problem. II,'' Phys. Rev.
{\bf 168} (5), 25 April 1968, p. 1415.}

\lref\hawkinghorowitz{S. W. Hawking and G. T. Horowitz, ``The Gravitational
Hamiltonian, Action, Entropy and Surface Terms," {\tt gr-qc/9501014}.}

\lref\horowitzsheinblatt{G. T. Horowitz and H. J. Sheinblatt,
``Tests of Cosmic Censorship in the Ernst Spacetime,'' {\tt gr-qc/9607027}.}

\lref\emparanreall{R. Emparan and H. S. Reall, ``Generalized Weyl Solutions,'' {\tt
hep-th/0110258}.}

\lref\tanteo{H. S. Tan and E. Teo, ``Multi-black hole solutions in five dimensions,''
{\tt hep-th/0306044}.}

\lref\gibbons{F. Dowker, J. P. Gauntlett, G. W. Gibbons, and G. T. Horowitz,
``The Decay of Magnetic Fields in Kaluza-Klein Theory,'' {\tt hep-th/9507143}.}

%\GutperleMB
\lref\GutperleMB{
M.~Gutperle and A.~Strominger,
``Fluxbranes in string theory,''
JHEP {\bf 0106}, 035 (2001)
[arXiv:hep-th/0104136].
%%CITATION = HEP-TH 0104136;%%
}

\lref\DavidsonDF{
A.~Davidson and E.~Gedalin,
``Finite Magnetic Flux Tube As A Black And White Dihole,''
Phys.\ Lett.\ B {\bf 339}, 304 (1994)
[arXiv:gr-qc/9408006],
%%CITATION = GR_QC 9408006,%%
}

%\LeblondDB
\lref\LeblondDB{
F.~Leblond and A.~W.~Peet,
``SD-brane gravity fields and rolling tachyons,''
JHEP {\bf 0304}, 048 (2003)
[arXiv:hep-th/0303035].
%%CITATION = HEP-TH 0303035;%%
}
%\KruczenskiAP
\lref\KruczenskiAP{
M.~Kruczenski, R.~C.~Myers and A.~W.~Peet,
%``Supergravity S-branes,''
JHEP {\bf 0205}, 039 (2002)
[arXiv:hep-th/0204144].
%%CITATION = HEP-TH 0204144;%%
}
%\ChenYQ
\lref\ChenYQ{
C.~M.~Chen, D.~V.~Gal'tsov and M.~Gutperle,
%``S-brane solutions in supergravity theories,''
Phys.\ Rev.\ D {\bf 66}, 024043 (2002)
[arXiv:hep-th/0204071].
%%CITATION = HEP-TH 0204071;%%
}
%\BuchelTJ
\lref\BuchelTJ{
A.~Buchel, P.~Langfelder and J.~Walcher,
%``Does the tachyon matter?,''
Annals Phys.\  {\bf 302}, 78 (2002)
[arXiv:hep-th/0207235].
%%CITATION = HEP-TH 0207235;%%
}
%\BurgessVU
\lref\BurgessVU{
C.~P.~Burgess, F.~Quevedo, S.~J.~Rey, G.~Tasinato and I.~Zavala,
%``Cosmological spacetimes from negative tension brane backgrounds,''
JHEP {\bf 0210}, 028 (2002)
[arXiv:hep-th/0207104].
%%CITATION = HEP-TH 0207104;%%
}

%\HeadrickYU
\lref\HeadrickYU{
M.~Headrick,
``Decay on C/Z(n): Exact supergravity solutions,''
arXiv:hep-th/0312213.
%%CITATION = HEP-TH 0312213,%%
}

\lref\bonnor{
W.~B.~Bonnor, ``Exact Solutions of the Einstein-Maxwell Equations,''
Zeitschrift f\"{u}r Physik {\bf 161}, p. 439 (1961).}

%\draft

\Title{\vbox{\baselineskip12pt\hbox{hep-th/0403050}\hbox{SLAC-PUB-10324}
\hbox{} }}{Non-Singular Solutions for S-branes}

\centerline{Gregory Jones\footnote{$^\dagger$}{Department of
Physics, Harvard University, Cambridge, MA 02138},
Alexander Maloney\footnote{$^*$} {SLAC and Department of Physics,
Stanford University, Stanford, CA 94309} and Andrew
Strominger$^\dagger$
}

\vskip .3in \centerline{\bf Abstract} {

Exact, non-singular, time-dependent solutions of Maxwell-Einstein
gravity with and without dilatons are constructed by double Wick
rotating a variety of static, axisymmetric solutions. This
procedure transforms arrays of charged or neutral black holes into
s-brane (spacelike brane) solutions, $i.e.$ extended, short-lived
spacelike defects. Along the way, new static solutions
corresponding to arrays of alternating-charge Reissner-Nordstrom
black holes, as well as their dilatonic generalizations, are
found. Their double Wick rotation yields s-brane solutions which
are periodic in imaginary time and potential large-N duals for the
creation/decay of unstable D-branes in string theory. } \vskip
.3in

\smallskip
\Date{}

\listtoc
\writetoc

%\vfill \eject

\newsec{Introduction}

Nearly a century after the discovery of general relativity, new
exact solutions continue to be found. In this paper we construct
several families of exact, {\it time}-{\it dependent} solutions
with and without electromagnetic fields and/or dilatons.  The
solutions are of the s-brane (spacelike brane) type \GutperleAI,
describing a shell of radiation coming in from infinity and
creating an unstable brane which subsequently decays. They are of
special interest because, in a variety of examples,  they are
singularity-free (outside horizons) and periodic in imaginary
time. In addition, they potentially provide large $N$ duals of
unstable D-brane creation/decay in string theory. A variety of
other s-brane gravity solutions can be found in
\refs{\LeblondDB\KruczenskiAP\ChenYQ\BuchelTJ-\BurgessVU}.

The s-brane solutions herein are generated by applying a trick of
``double Wick rotation'' to static solutions \bonnor.  Typically, a Wick
rotation $z \to it$ is applied to a Killing direction
parameterized by a coordinate $z$ to obtain a static, Lorentzian
solution from a Euclidean one.  The Killing symmetry implies that
the metric is $z$-independent and hence remains real under $z \to
it$. However, this reality is guaranteed by the weaker condition of
a discrete symmetry $z \to -z$, implying that the metric contains
no odd powers of $z$, as well as a suitable condition on the
electromagnetic field if present. In this more general context Wick rotation
leads to time-dependent solutions. Of course there is no guarantee
that the resulting solutions will be singularity-free in general.

We apply this trick to several interesting static solutions in the
literature, including the recently understood black dihole
solution \EmparanAU\ representing a pair of oppositely charged
black holes and the Israel-Khan infinite line array of neutral
black holes \refs{\IsraelKhan, \myers}.  They are Wick rotated
twice to obtain a new Lorentzian solution: first along the static
time direction \eqn\wick{\tau \to iy} to obtain a Euclidean
solution and then again along the spacelike discrete symmetry
direction (the axis of the black holes) \eqn\wickk{z\to it}
 to obtain a new Lorentzian
solution. Interestingly enough, while the original black dihole
solution has conical ``strut'' singularities, these are moved off
into the complex plane and do not appear in the singularity-free
s-brane solution.

Neither of the solutions so obtained quite corresponds to the type
of s-brane most naturally arising in string theory.\foot{The black
dihole s-brane corresponds to generalized sD-brane configurations
of the type described in \GaiottoRM.} These are related by double
Wick rotation to a (previously unknown) solution describing a
static linear array of alternating-charge black holes. The
linearized versions of these solutions were found in \MaloneyCK\
using Sen's open string boundary state \SenNU\ for the
creation/decay of an unstable D-brane. Here we construct an exact
non-linear solution describing just such an array, and find that
(unlike the dihole case) it is free of strut singularities.  The
double Wick rotation of this solution is then the non-linear
generalization of the linearized s-brane solutions in \MaloneyCK.
A salient property of these solutions, desired for a connection to
string theory and distinguishing them from previous s-brane gravity
solutions, is that they are periodic in imaginary time. They bear
roughly the same relation to the unstable D-brane creation/decay,
that the black $p$-brane solutions \HorowitzCD\ bear to ordinary
stable D-branes. Hence they may be relevant for a time-dependent
large $N$ duality.

This paper is organized as follows. Section 2 begins with a brief
review of the Weyl formalism for axisymmetric solutions of the
Einstein-Maxwell equations. Emparan's black dihole solution is
then described, and Wick-rotated to a smooth s-brane solution. The
s-charge is computed and found to be nonzero. Dilatonic
generalizations are described. In section 3 we find that double
Wick rotation of the Israel-Khan solution for a neutral black
hole pair (with strut singularities) leads to a smooth s-brane
solution. We also construct the double Wick rotation of the
(strut-free) Myers neutral black hole array solution to an s-brane
solution. In section 4, following a suggestion of Gibbons \gary,
we construct a static solution corresponding to an alternating
array of KK monopoles and antimonopoles, containing a dilaton from
dimensional reduction. The dilaton is removed using the general
method of \EmparanBB, yielding a static line array of
alternating-charge Reissner-Nordstr\o m solutions of the pure
Einstein-Maxwell theory.  We perform the double Wick rotation
\wick, \wickk\
and study various properties of the resulting s-brane solution.
Finally, in section 5 we discuss
the relevance of these gravity solutions to string theory.

\newsec{S-Branes from Black Diholes}

We will start by reviewing the general construction of static, axisymmetric
solutions of Einstein-Maxwell theory in Section 2.1.
We will then proceed to describe
%We will start by describing
the simplest examples of double Wick rotated
solutions.  These are found by starting with static solutions describing
a pair of oppositely charged, extremal black holes.
As we will show in
Section 2.2, the double Wick rotation
procedure \wick, \wickk\ transforms this into an
exact, time dependent solution that is everywhere smooth and non-singular.

These solutions for extremally charged black holes have
a direct string theory interpretation in terms of
D-branes.  In particular, dilatonic generalizations
of these solutions can be lifted to give s6-brane solutions of IIA
supergravity, as is described in Section 2.3.

%Before tackling the full solutions,
%

\subsec{Time Dependent Axisymmetric Einstein-Maxwell Solutions}

A static, axisymmetric solution of Einstein-Maxwell theory
has a metric of the form
\eqn\asd{\eqalign{
ds^2 &= -e^{X}d\t^2 + e^{-X}(\rho^2 d\varphi^2+e^{2\gamma} (d\rho^2 + dz^2)),\cr
}}
and an electrostatic potential
\eqn\asd{\eqalign{
A_\mu dx^\mu &= A d\t
.}}
Here $X$, $\gamma$ and $A$ are functions of $\rho$ and $z$ only.
We will use units $G=1$.
Upon double Wick rotation \wick, \wickk\ this becomes a
time dependent solutions of the form
\eqn\asd{\eqalign{
ds^2 &= e^{X}dy^2 + e^{-X}(\rho^2 d\varphi^2+e^{2\gamma} (d\rho^2 - dt^2)),\cr
A_\mu dx^\mu &= i A dy
.}}
In order for the metric and gauge potential to be real, we see that
$A$ must be an antisymmetric
function of $z$, whereas $e^X$ and $e^{2\gamma}$ must be symmetric.

The sourceless Maxwell equation is
\eqn\max{
\ddot A + A'' + {1\over \rho} A' = \dot A \dot X + A' X'
,}
where prime denotes $\partial/\partial\rho$ and dot denotes $\partial/\partial z$.
Einstein's equation implies that
\eqn\eeq{\eqalign{
\ddot X + X'' + {1\over \rho} X' &= 2 e^{-X} (\dot A^2 + A'^2)
}}
and
\eqn\gis{\eqalign{
{4\over \rho} \dot \gamma &= 2 X' \dot X  - 8 e^{-X} \dot A A' \cr
{4\over \rho} \gamma' &= X'^2 - \dot X ^2 + 4 e^{-X} (\dot A^2 - A'^2)
.}}
%All additional components of the field equations follow from these.
Equation \gis\ determines $\gamma$ in terms
of $A$ and $X$.  One might worry that these overconstrain $\gamma$,
but in fact
this system is integrable, as can be seen by testing the
equality of mixed partials in \gis.  One important feature of
\gis\ is that along the $z$-axis (i.e. for $\rho=0$), $\dot\gamma$
vanishes in the absence of sources.  Thus
any conical singularities that are present along the
$z$-axis will be constant.  As we will see below,
this fact enables us to get rid of conical
singularities altogether in certain double Wick rotated geometries.

For future reference, we will note one important symmetry of
\eeq -\gis.  With $A=0$ (pure Einstein
theory), they have the symmetry
\eqn\ttmsym{X\to -X+2\log\rho+{\rm constant},\qquad
\gamma\to\gamma-X+\log\rho+{\rm constant}.}
As we will see in section 4,
this transformation of $X$ corresponds to turning off black holes where
they had existed on the $z$-axis, and turning on black holes where they had not
existed.  The accompanying transformation of $\gamma$ is required by
Einstein's equations.

It remains to solve for $X$ and $A$ using \max\ and \eeq.
As was first noted by Weyl \Weylpaper\EmparanWK, these equations are particularly
simply when written in terms of
a new three dimensional space with coordinates
$(\varphi, \rho, z)$ and flat metric
\eqn\asd{
d{\hat s}^2 = d\rho^2 + dz^2 + \rho^2 d\varphi^2 .
}
The equations of motion are
\eqn\eomm{
\nabla^2 X = 2e^{-X} |\nabla A|^2,~~~~~
\nabla^2 A = \nabla A \cdot \nabla X, }
where $\nabla$ is the covariant derivative with respect to $d{\hat s}^2$.

Although beguilingly simple, these equations are difficult to solve
analytically when $A\ne 0$.  The dihole solution of the next section
is one of few such exact solutions that admits a sensible
black hole interpretation.

\subsec{Double Wick Rotation of the Black Dihole}

The black dihole solution
\eqn\bdi{\eqalign{
ds^2 &= - e^X d\tau^2 + e^{-X} (\rho^2 d\varphi^2 + e^{2\gamma} (d\rho^2 + dz^2)),~~~~~
A_\mu dx^\mu = A d\tau \cr
e^X &= \left[ { (R_+ + R_-)^2 - 4m^2 - {k^2\over m^2 + k^2} (R_+-R_-)^2
\over (R_+ + R_- + 2m)^2 - {k^2\over m^2 + k^2} (R_+-R_-)^2}\right]^2 \cr
e^{2\gamma} &=\left[ { (R_+ + R_-)^2 - 4m^2 - {k^2\over m^2 + k^2} (R_+-R_-)^2
\over 4 R_+ R_-}\right]^4  \cr
 A &= -{4mk \over \sqrt{m^2 + k^2}} {R_+ - R_-
\over (R_+ + R_- + 2m)^2 - {k^2\over m^2 + k^2} (R_+-R_-)^2} \cr
R_\pm &= \sqrt{ \rho^2 + (z \pm \sqrt{ m^2 + k^2 })^2}
}}
has been studied by many authors \refs{\EmparanAU,\EmparanBB,\TeoUG}.
We will summarize here only a few
salient features.
The solution \bdi\
describes a static configuration of
two extremal black holes of equal mass
and opposite charge.  The solution is written in terms of parameters
$m$ and $k$; $2m$ is the total mass of the system and the black
holes are located at $z=\pm \sqrt{m^2+k^2}$, $\rho=0$.
The black holes are extremal in
the sense that their horizons are degenerate.
The metric has conical singularities extending along the $z$-axis, as may
be seen from
the limiting form of the metric in the $(\rho,\varphi)$ plane at small $\rho$ and $-\sqrt{m^2+k^2}<z<\sqrt{m^2+k^2}$
\eqn\asd{e^{-X}\left(\big({k^2\over m^2+k^2}\big)^4d\rho^2
+\rho^2d\varphi^2\right).}
The standard identification $\varphi\simeq\varphi+2\pi$ leads to a
conical singularity
on the interval $-\sqrt{m^2+k^2}<z<\sqrt{m^2+k^2}$ extending between the two
black holes -- this may be thought of as a strut balancing the
attractive gravitational and electric forces between the
two black holes \EmparanAU.
This singularity passes through the origin $z=0$, so
would lead to a singular solution after the double Wick rotation \wick, \wickk.
We will avoid this problem by noting that the equation of motion \gis\ fixes
$\gamma$ only up to a constant of integration.
Thus we are free to subtract the constant
\eqn\gsub{\gamma\to\gamma-2\log{k^2\over m^2+k^2}.}
With this choice, conical deficits are
located instead on the semi-infinite intervals $z<-\sqrt{m^2+k^2}$ and $z>\sqrt{m^2+k^2}$.
This corresponds to hanging the black holes from infinity by cosmic strings
rather than inserting a strut between them.  % to keep them apart.
The subtraction \gsub\ is equivalent to
adopting the non-standard periodicity for $\varphi$
\eqn\phip{
\varphi\simeq \varphi+2\pi\left({k^2\over m^2+k^2}\right)^2.}

Applying the double Wick rotation \wick,
\wickk\ gives a time-dependent solution,
which we will write as
\eqn\asbd{\eqalign{
ds^2 &= e^{-X}(\rho^2 d\varphi^2+e^{2\gamma} (d\rho^2 -dt^2))+e^{X}dy^2,~~~~~
A_\mu dx^\mu = A\,dy
}}
where
\eqn\sbd{\eqalign{
%e^{X}&=\left[{m^2(x^2-1) + k^2 |R|^2\over (k^2 + m^2) (x+m)^2 + k^2 y^2 }
%\right]^2  \cr
e^X &= \left[ {(\Re R)^2 - m^2 + {k^2\over m^2 + k^2} (\Im R)^2
\over (\Re R + m)^2 + {k^2\over m^2 + k^2} (\Im R)^2 }
\right]^2\cr
%e^{2\gamma}&=\left[k^2 + m^2{(x^2-1)\over|R|^2}\right]^4  \cr
e^{2\gamma} &= \left({m^2+k^2\over k^2}\right)^4\left[ {(\Re R)^2 - m^2 + {k^2\over m^2 + k^2} (\Im R)^2
\over |R|^2}\right]^4\cr
%A &= {2mk\sqrt{m^2+k^2}} {y\over (k^2 + m^2) (x+m)^2 + k^2 y^2 }
A &= {2mk\sqrt{m^2+k^2}} {\Im R\over (k^2 + m^2) (\Re R+m)^2 + k^2 (\Im R))^2 }
.}}
Here we've defined the complex distance to one of
the sources\foot{This definition $R=\sqrt{\rho^2+m^2+k^2-t^2-2it\sqrt{m^2+k^2}}$
involves a branch prescription; we will analytically continue from $t=0$
and note that the radicand traces a parabola which does not intersect
the standard branch cut, on the negative real axis.} \eqn\ris{\eqalign{
R &= \sqrt{\rho^2 -(t+i\sqrt{m^2+k^2})^2}
.}}
It is straightforward to verify that the metric and gauge potential
are smooth and non-singular for all real values of $t$ and $\rho$.
%of this.

This solution has translational Killing symmetry $\p_y$ in addition to
the usual rotational vector
$\p_\varphi$, so resembles the creation and subsequent decay of a
1-brane whose spatial direction extends in the $y$ direction.
Suppressing the $y$ direction, the solution describes a spherical wave of
gravitational and electromagnetic flux that is localized around the
light cone $\rho^2=t^2$.$\,$\foot{This solution has a translational symmetry in the $y$ direction, so
may naturally be thought of as a
spherical wave in the other 2+1 dimensions.  It therefore resembles the
2+1 dimensional supergravity solutions of \HeadrickYU\ describing the decay of
${\bf C} / {\bf Z}_n$ singularities.}
The solution is asymptotically flat either at
large radius or in the far past or future.
The gravitational potential $g_{tt}$, the warp factor
$e^{X}$ and the potential $A$ are shown in Figure 1 for typical
values of $k$ and $m$.

\triplefig{
%FIGURE
Various components of the black dihole S-brane solution:
(a) the gravitational potential $g_{tt}$, (b) the warp factor $e^{X}$, and
(c) the gauge potential $A$.  All components are non-singular.
}{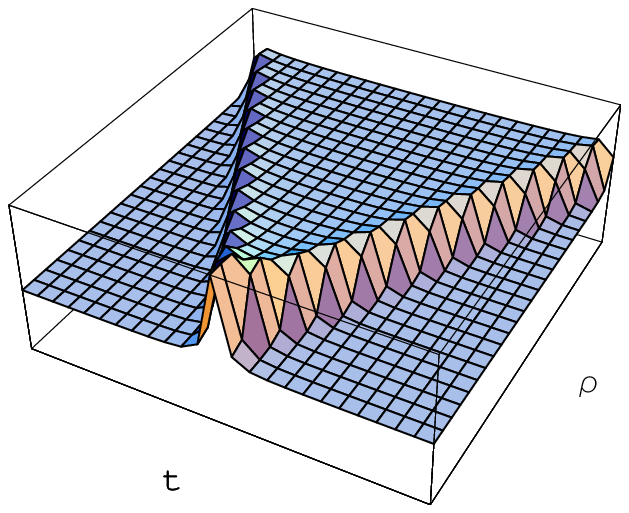}{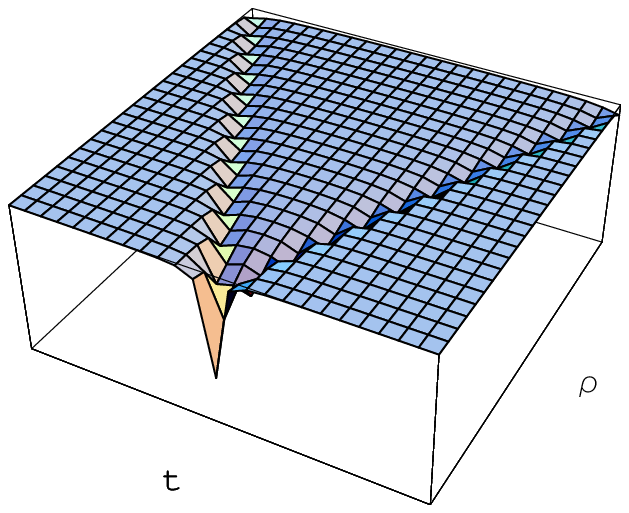}{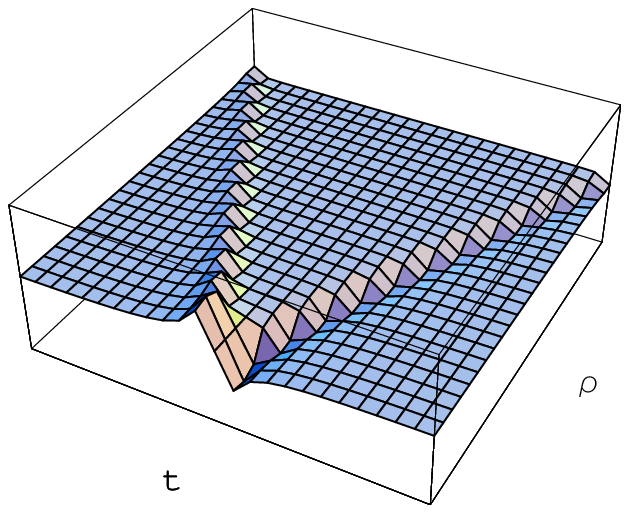}{0.0truein}

%greg's figures are
%(a) e^X di1.eps
%(b) F_{ty} di2.eps

Outside the light cone $\rho^2=t^2$, the fields asymptote to Minkowski space.
As one crosses the light cone the
gravitational potential $g_{tt}$ and the gauge field $A$ jump,
indicating that an observer feels gravitational and electromagnetic forces,
as is typically the case for s-brane solutions. %as she crosses the
There
is a nonzero
conserved s-charge found by integrating ${1\over 4\pi}*F$ along a complete
transverse spacelike slice (which we can take to be the $\rho\varphi$
plane).
It is sufficient to compute $F=\partial A/\partial t\,dt\wedge dy+\cdots$
and evaluate the charge at time $t=0$.  We find
\eqn\qis{\eqalign{
Q_s&={1\over 4\pi} \int {(\sqrt{\rho^2+m^2+k^2}-m)^2 2mk\over
(\rho^2+k^2)^2(\rho^2+m^2+k^2)^{1/2}}\,\rho d\rho\wedge d\varphi \cr
&={m\over k}(\sqrt{m^2+k^2}+m).}}
This is the same as the charge magnitude $Q$ of either extremal black hole
in the original black dihole geometry \bdi.  To see that this
must be
the case, consider computing $Q$ by integrating ${1\over 4\pi}*F$ on
a smooth topological sphere $S$ surrounding the black hole at
$z=\sqrt{m^2+k^2}$.
Now deform and enlarge $S$ so that it is the boundary of a half-ball; $S$
comprises a disk in the $z=0$ plane and a half-sphere centered at $z=0$, $\rho=0$.
Taking the limit where the radius of the disk and half-sphere goes to infinity,
the disk becomes the $(\rho,\varphi)$ plane and the flux through the half-sphere
goes to zero.  The quantity $F_{tz}|_{z=0}dt\wedge dz$ is unaffected
(up to a conventional sign) by the double Wick rotation, and the result
\qis\ follows.

This geometry has an interesting limit where the parameter
$m$ is large compared to $k$, $\rho$, and $t$.  Specifically, we take
$m\to m/\lambda$, $(k,\rho,t)\to(k,\rho,t)/\lambda^2$, and
$y\to\lambda^2 y$ where $\lambda\to\infty$.
%(We keep
%$\phi\simeq \phi+2\pi$ and conical singularities are always
%absent.)
The physical significance of this limit will be
discussed shortly.  Expanding the functions in \asbd -\sbd, we find
in this limit
\eqn\mel{\eqalign{ ds^2 &= \Big({k^2 + \rho^2\over
4m^2}\Big)^2\Big[ dy^2 +\big({4m^2\over k^2}\big)^4 (-dt^2 +
d\rho^2)\Big] + \Big({4m^2\over k^2 + \rho^2}\Big)^2\rho^2
d\varphi^2\cr A &= -(kt/2m^2) dy, ~~~~~ F= -(k/2m^2) dt\wedge dy.}}
This is the Melvin solution describing a constant electric field
pointing in the $y$ direction, with physical strength ${\vec
E}\cdot\hat{y} = {k\over 2m^2}\big({k^2\over k^2+\rho^2}\big)^2$.
%the decay is 1/\rho^4
These solutions have been well studied in the context of string theory
\GutperleMB\ -- they are typically written in terms of the dual
field strength
%The magnetic dual of this solution has the more familiar form
\eqn\asd{
*F = -{8km^2\rho d\rho\wedge d\varphi\over (k^2 + \rho^2)^2},
}
which represents a constant magnetic field.

The appearance of the Melvin solution has a natural physical
interpretation. Consider the geometry before the double Wick
rotation. The black holes are located at
$z=\pm\sqrt{m^2+k^2}$, which becomes $z=\pm\lambda m$ in our limit.
As $\lambda\to\infty$
the black holes are pulled off to
infinity and we are left with the electric flux tube -- which is
described by the Melvin universe -- running between them.
This relation between static black diholes and the
Melvin universe was studied in \EmparanGM.
The same idea applies to the double Wick rotated geometry, since
the Melvin solution \mel\ is preserved under double Wick
rotation up to a rescaling of $t$.
Here, as
the sources recede away along the imaginary $t$ axis the solution
becomes static.

From this discussion, we see that for $k\ll m$, the solution \sbd\
can be thought of as the creation and subsequent decay of an
electric  Melvin flux tube located near $\rho=0$.

\subsec{Adding a Dilaton}

The dihole solutions described above
have straightforward dilatonic generalizations, which
were studied by \refs{\EmparanBB,\EmparanGM,\DavidsonDF,\Liang}.
The details of the double
Wick rotation are essentially the same as in the non-dilatonic case,
so we will just state the results.

The s-brane solution of Einstein-Maxwell-Dilaton theory with
dilaton coupling $\alpha$ and extremal electric sources is
\eqn\sbds{\eqalign{ ds^2 &=
e^{-X/(1+\alpha^2)}(\rho^2 d\varphi^2+e^{2\gamma/(1+\alpha^2)}
(d\rho^2 -dt^2)) +e^{X/(1+\alpha^2)}dy^2\cr A_\mu dx^\mu &=
{1\over \sqrt{1+\alpha^2}} A\,dy,~~~~~ e^{2\phi}=e^{\alpha X /
(1+\alpha^2)}, }} where $X$, $\gamma$ and $A$ are as in \sbd.
This
solution is smooth and non-singular for all real $\rho$ and $t$.
The structure of this solution is very similar to the
configuration described above. These dilatonic solutions can be
lifted to give solutions of IIA string theory.  In this case the
closed string configurations are sourced by Euclidean D6-branes,
and are related to the creation and subsequent decay of an
unstable D7-brane.

As one might expect, the $m\to\infty$
limit of this solution gives the electric Melvin
solution of string theory.
This is the electric dual of the fluxbrane solutions written down
by \GutperleMB.

%We will make one final observation, which will be important when
%we consider more complicated dilatonic solutions in section 4.
%The ordinary extremal Reissner-Nordstr\o m black hole \EmparanBB\ is
%represented in Weyl space
%by a point at which $e^X$ vanishes to second
%order.  %, as is typical at the horizon of an extremal black hole.
%Its
%dilatonic generalization (which is \sbds\ before double Wick rotation)
%for $\alpha=\sqrt{3}$
%then gives $-g_{\tau\tau}=e^{X/4}$, which has a zero of order one-half instead of
%order two.

\newsec{Neutral S-Brane Solutions}

In this section we study the general time-dependent solutions
found by Wick rotating an axisymmetric collection of sources on the
$z$-axis with no charges.  Unlike the solutions of the previous section, these
configurations
are not sourced by extremal objects on the $z$-axis, and hence do not
have a simple string theory
interpretation in terms of Euclidean D-branes.  Nevertheless,
they are interesting new, non-singular, time-dependent solutions of general relativity and
string theory.

We will first describe the Wick rotation of a pair of
neutral black holes -- the details of this construction are similar to that
of the dihole in the previous section.  Next, we apply this
procedure to the periodic array of neutral black holes described by Myers
\myers.  This yields an s-brane type solution that is explicitly periodic in
imaginary time.

\subsec{Double Wick Rotated Black Hole Pair}

When there are no electromagnetic fields ($A=0$), the equations of
motion \eomm\ simplify considerably.
Axisymmetric solutions
to the sourceless Einstein equations are found by solving Laplace's
equation in three dimensions,
\eqn\xeom{\nabla^2 X =0.}
The general solution is found by specifying a density of
sources $b(z)$ distributed along the $z$-axis. In this case we can
immediately write down the solution
\eqn\densityint{
X(\rho,z)= -2\int_{-\infty}^\infty dz' {b(z')\over \sqrt{\rho^2+(z-z')^2}}}
and proceed to solve for $\gamma$ using $\gis$.
To avoid naked singularities,
the linear density $b(z)$ must typically equal $1/2$ on some line segments
and be zero elsewhere \EmparanWK.  A simple solution is given by placing a
rod source of length $2m$ and density $b=1/2$ on the $z$-axis.  This gives
\eqn\schwarz{\eqalign{
e^X&={R_+ +r_+ -2m\over R_+ +r_+ +2m},~~~~~
e^{2\gamma} = {(R_+ + r_+)^2 -4m^2 \over 4 R_+ r_+}\cr
R_+ &=\sqrt{\rho^2 +(z+m)^2}, ~~~~~ r_+ =\sqrt{\rho^2 +(z-m)^2}
.}}
In fact, this is precisely the Schwarzschild solution with mass $m$.
The coordinate transformation between \schwarz\ and the usual
Schwarzschild coordinates may be found in \EmparanBB.

Laplace's equation is linear, so we can easily construct from this
multi-black hole
solutions by superposing copies of the Schwarzschild solution \schwarz.
These are known as the Israel-Khan solutions \IsraelKhan.
For example, the solution for two separated black holes is
\eqn\ik{\eqalign{
e^X&={(R_++r_+ - 2m)(R_-+r_--2m)\over (R_++r_+ + 2m)(R_-+r_-+2m)}\cr
%e^{2\gamma} &= XXX  \cr
R_\pm &=\sqrt{\rho^2 +(z\pm(k+m))^2}, ~~~~~
r_\pm =\sqrt{\rho^2 +(z\pm(k-m))^2}
.}}
Here
the two black holes sources are located
on the intervals $-k-m<z<-k+m$ and $k-m<z< k+m$.
In this case the double Wick-rotated solution is
\eqn\ikk{\eqalign{
ds^2 &= e^{-X} (\rho^2 d\varphi^2 + e^{2\gamma}(d\rho^2 - dt^2)) + e^{X} dy^2 \cr
e^{X}&=\left|R+r - 2m\over R+r + 2m\right|^2\cr
R &=\sqrt{\rho^2 -(t+ i(k+m))^2}, ~~~~~
r =\sqrt{\rho^2 -(t+i(k-m))^2}
.}}
The expression for $e^{2\gamma}$ can be found explicitly \myers:
\eqn\asd{
e^{2\gamma}=\Big({k^2\over k^2-m^2}\Big)^2\,{(|R|^2+\rho^2-t^2-(k+m)^2)(|r|^2+\rho^2-t^2-(k-m)^2)\over
|R\overline{r}-t^2-2mit-k^2+m^2+\rho^2|^2}
}
%\eqn\asd{\eqalign{
%e^{2\gamma}&=\left( {k^2\over m^2+k^2}\right)^2\left| { (\rho^2 - %(t-i(k+m))(t-i(k-m))+Rr)\over
%2Rr(\rho^2 - |t+i(k+m)|^2+R\br)}  \right|^2 \ \times \cr
%&~~~~~~~~~~(\rho^2-t^2-(k-m)^2 + |r|^2)(\rho^2-t^2-(k+m)^2 + |R|^2)
%.}}
As in the dihole case, we have chosen the prefactor
so that the Wick rotated solution is smooth and non-singular for
all real $t$, $y$ and $\rho$.
With this choice of $\gamma$, the original solution \ik\
has conical singularities located on the semi-infinite intervals
$z<-m-k$, $z>m+k$ --- these are interpreted as
cosmic strings that balance
the attractive force between the black holes.

\subsec{Double Wick-Rotated Black Hole Array}

We can apply this procedure to find the solution for an infinite
array of Schwarzschild black holes, with rod-centers at $z=(2p+1)k$
for integers $p$, and mass $m$:
\eqn\asdz{\eqalign{
 e^X&=\prod_{p=-\infty}^\infty {R_p+r_p-2m\over R_p+r_p+2m}\cr
e^{2\gamma} &=
\prod_{p,\,q=-\infty}^\infty {r_q R_p+(z-(2q+1)k-m)(z-(2p+1)k+m)+\rho^2\over
r_q r_p+(z-(2q+1)k-m)(z-(2p+1)k-m)+\rho^2} \times \cr
&~~~~~~~~~~   {R_q r_p+(z-(2q+1)k+m)(z-(2p+1)k-m)+\rho^2\over
R_q R_p+(z-(2q+1)k+m)(z-(2p+1)k+m)+\rho^2} \cr
R_p &=\sqrt{\rho^2 +(z-(2p+1)k+m)^2}, ~~~~~ r_p =\sqrt{\rho^2
+(z-(2p+1)k-m)^2} .
}}
%\eqn\asdz{\eqalign{
% e^X&=\Pi_{N} {R_{N}+r_{N}-2m\over R_{N}+r_{N}+2m}\cr
%e^{2\gamma} &=
%\Pi_{NM} {(R_N r_M + (z+(Nk+m))(z+(Mk-m)) +\rho^2) \over
%         (R_N R_M + (z+(Nk+m))(z+(Mk+m)) +\rho^2)} \times \cr
%&~~~~~~~~~~   {(r_N R_M + (z+(Nk-m))(z+(Mk+m)) +\rho^2) \over
%         (r_N r_M +(z+(Nk-m))(z+(Mk-m)) +\rho^2)} \cr
%R_{N} &=\sqrt{\rho^2 +(z+(Nk+m))^2}, ~~~~~ r_N =\sqrt{\rho^2
%+(z+(Nk-m))^2} .}} Here $N$ runs from $-\infty$ to $\infty$.
These sums formally diverge, but may be regularized using a
procedure employed by Myers \myers.  For solutions of pure
gravity, the equations of motion \gis\ and \xeom\ determine
$\gamma$ and $X$ only up to constants of integration, so we may
subtract off the divergent constants \eqn\asds{\eqalign{
X_{\rm divergent} &= \sum_{p=-\infty}^\infty \ln \left({1-m/|2p+1|k\over
1+m/|2p+1|k}\right) \cr
\gamma_{\rm divergent} &= \sum_{p\geq0,\,q\leq -1}
\ln\left({1-m^2/(p-q)^2k^2}\right) .}} Any further finite
subtraction in $X$ is up to us; this is the usual ambiguity in the
potential from a linearly extended source.  However, the
subtraction in $\gamma$ has been chosen such that the solution has no
conical singularities and is smooth and non-singular. The
ability to eliminate all conical singularities in the geometry
is not surprising, since for this
configuration the forces between the black holes are balanced.
The double Wick rotated solution is then
\eqn\asdz{\eqalign{
e^X&=e^{-X_{divergent}} \prod_{p=-\infty}^\infty {R_p+r_p-2m\over R_p+r_p+2m}\cr
e^{2\gamma} &= e^{-2\gamma_{divergent}}
\prod_{p,q=-\infty}^\infty {r_q R_p - (t+i(2q+1)k+im)(t+i(2p+1)k-im)+\rho^2 \over
r_q r_p -(t+i(2q+1)k+im)(t+i(2p+1)k+im)+\rho^2} \times \cr
&~~~~~~~~~~   {R_q r_p-(t+i(2q+1)k-im)(t+i(2p+1)k+im)+\rho^2 \over
R_q R_p -(t+i(2q+1)k-im)(t+i(2p+1)k-im)+\rho^2} \cr
R_p &=\sqrt{\rho^2 -(t+i(2p+1)k+im)^2}, ~~~~~ r_p =\sqrt{\rho^2
-(t+i(2p+1)k-im)^2} .}}
%\eqn\asdz{\eqalign{
% e^X&=\Pi_{N} {R_{N}+r_{N}-2m\over R_{N}+r_{N}+2m}\cr
%e^{2\gamma} &=
%\Pi_{NM} {(R_N r_M - (t-i(Nk+m))(t-i(Mk-m)) +\rho^2) \over
%         (R_N R_M - (t-i(Nk+m))(t-i(Mk+m)) +\rho^2)} \times \cr
%&~~~~~~~~~~   {(r_N R_M - (t-i(Nk-m))(t-i(Mk+m)) +\rho^2) \over
%         (r_N r_M - (t-i(Nk-m))(t-i(Mk-m)) +\rho^2)} \cr
%R_{N} &=\sqrt{\rho^2 -(t-i(Nk+m))^2}, ~~~~~ r_N =\sqrt{\rho^2
%-(t-i(Nk-m))^2} .}}

\newsec{Alternating Charge S-brane Array}

In this section we find a new  solution of
Einstein-Maxwell theory in 3+1 dimensions (with or without a dilaton)
which
corresponds to a non-supersymmetric, infinite array of
alternating-charge extremal black holes. We then double Wick
rotate this solution, yielding an exact solution for s-branes of
the type encountered in string theory \MaloneyCK. In particular,
these solutions are periodic in imaginary time. We thank Gary
Gibbons for suggestions leading to the construction of this
section.

\subsec{Static Alternating-Charge Extremal Black Hole Array}

We begin with the Israel-Khan solution with sources of length $k$
centered at $z=(2p+1)k$ given in \asdz.
Analytically continuing the time coordinate $\tau\to ix^5$ gives the
Euclidean solution \eqn\asd{ ds^2=e^X
(dx^5)^2+e^{-X}(e^{2\gamma}(d\rho^2+dz^2)+\rho^2d\varphi^2).} In
analogy with Euclidean Schwarzschild, we anticipate that
identifying $x^5$ on a thermal circle $x^5\simeq x^5+2\pi R$ for
some $R$  will yield a non-singular geometry.  The black hole
horizons are replaced by 2-surfaces where $\partial/\partial x^5$
vanishes; these surfaces are termed `bolts' \GibbonsHawking, and
are along the rods at $\rho=0$ where $e^X$ vanishes.  We have
\eqn\tio{R=\lim_{\rho\to 0}\rho e^\gamma/e^X.}
We see that $R$ depends on the infinite constant we subtracted
from $X$; we can subtract from $X$ and scale $R$ such that $Re^X$ remains
constant, and maintain nonsingular bolts.
%We find that
%the geometry would be
%regular if we identify $x^5$ on a circle of size
%\eqn\tio{R=4m.}

However we wish to do something a bit different---namely  a
twisted KK compactification, as in \DowkerGB. We first add a flat
time direction (as in the construction of the KK monopole
\refs{\GrossPerry,\Sorkin}) to get \eqn\asdb{
ds^2=-d\tau^2+e^{-X}(e^{2\gamma}(d\rho^2+dz^2)+\rho^2d\varphi^2)+e^X(dx^5)^2.}
%Instead of $\varphi\simeq\varphi+2\pi$ and $x^5\simeq x^5+2\pi R$
%as above, we do the twisted identification
%\eqn\asdc{(\varphi,x^5)\simeq(\varphi+2\pi n,x^5+2\pi Rn),
%~~~~{\rm for~~ any ~~integer}~~ n.}
%greg's note: the above equation is not actually a twisted identification,
%since it is discrete and those points were identified anyway
Consider the Killing vector $K =R\partial_5+\partial_\varphi$; its
fixed points are the north and south poles of each horizon. Such
isolated fixed points are known as nuts or antinuts, depending on
the relative orientation of the $SO(2)\times SO(2)$ rotation
induced by the Killing vector in the tangent space at the
nut/antinut \GibbonsHawking. The north pole will be a (self-dual)
nut, and the south pole an (anti-self-dual) antinut
\GibbonsHawking; after Kaluza-Klein reduction, these will become a
magnetic monopole and antimonopole, respectively.
\foot{This observation is related to the fact \EmparanAU\ 
that the black dihole solution with appropriate dilaton coupling
is related to a Schwarzchild instanton.}

An equivalent procedure is to change coordinates to
$\tilde\varphi=\varphi-x^5/R$ and then reduce along orbits of
$\partial_5$ with $\tilde \varphi$ held fixed. This turns out to be
notationally simpler. In the new coordinate, the metric is
\eqn\asdd{ds^2=-d\tau^2+e^{-X}(e^{2\gamma}(d\rho^2+dz^2)+\rho^2d\tilde\varphi^2)
+2e^{-X}{\rho^2 \over R}d\tilde\varphi
dx^5+(e^X+{e^{-X}\rho^2\over R^2})(dx^5)^2.}

Now, perform KK compactification  with $\partial/\partial x^5$ as
the Killing vector.  Using \eqn\asde{ d\hat
s^2=e^{-4\phi/\sqrt{3}}(dx^5+2A_\mu
dx^\mu)^2+e^{2\phi/\sqrt{3}}g_{\mu\nu}dx^\mu dx^\nu,} which
yields the Einstein frame action
\eqn\rphifaction{ S={1\over 16\pi G_4}\int d^4x
\sqrt{-g}(R-2(\nabla\phi)^2-e^{-2\sqrt{3}\phi}F^2),} we get
\eqn\bigsolution{\eqalign{ ds^2&=(e^X+e^{-X}{\rho^2\over
R^2})^{1/2}\left(-d\tau^2+e^{-X}e^{2\gamma}(d\rho^2+dz^2)+
{\rho^2\over e^X+e^{-X}\rho^2/R^2}d\tilde\varphi^2\right)\cr A_\mu
dx^\mu &={\rho^2R\over 2(\rho^2+R^2e^{2X})}d\tilde\varphi\cr
e^{-4\phi/\sqrt{3}}&=e^X+e^{-X}{\rho^2 \over R^2}. } }

We wish to look for possible singularities on the symmetry axis
$\rho= 0$. In the sections of the $z$-axis where there had been no
rods, $e^X\to({\rm nonzero})$ as $\rho\to 0$, and the relevant
part of the metric is
$$e^{-X}e^{2\gamma}d\rho^2+{\rho^2\over e^X+e^{-X}\rho^2/R^2}d\tilde\varphi^2
\quad\to\quad
e^{-X}\Big(e^{2\gamma}d\rho^2+\rho^2d\tilde\varphi^2\Big)$$ so the
absence of singularities in the Israel-Khan solution guarantees
their absence here. On the other hand, in the sections where there
had been rods, $e^X\sim\rho^2$ and the metric looks like
\eqn\wsd{e^{-X}e^{2\gamma}d\rho^2+{\rho^2\over
e^X+e^{-X}\rho^2/R^2}d\tilde\varphi^2 \quad\to\quad
e^{-X}e^{2\gamma}d\rho^2+e^X R^2d\tilde\varphi^2.}
Now we see that the same
choice of $x^5$ periodicity \tio\
which makes the Euclideanized
Israel-Khan solution regular at the horizons,
ensures that our twisted KK-reduced solution
is regular along the rod locations.

Although the 5D geometry is smooth, there are singularities in the
4D description at the rod endpoints. These are locally as
considered in \refs{\GrossPerry,\Sorkin} and correspond to KK
monopoles or anti-monopoles.  For $z$ at endpoints of a rod,
$e^X\sim \rho$ as $\rho\to 0$, whereas for $z$ in the interior of
a rod, $e^X\sim\rho^2$ as $\rho\to 0$. From this we compute that
$g_{tt}\to 0$ as $\rho^{1/2}$ at the endpoints while
$g_{tt}\to{\rm\ (nonzero)}$ on the interior of the rods.  This
signifies the presence of extremal $\alpha=\sqrt{3}$ dilatonic
black holes at the endpoints of the rods.  There
are Dirac strings running along the rods, joining each
monopole-antimonopole pair. The monopole charge can be computed
from the Dirac string: we simply place a sphere around the monopole with a
neighborhood of the Dirac string deleted. This yields
\eqn\asd{Q={1\over 4\pi}\int_{\rm deleted\ sphere} dA = {1\over
4\pi}\oint_{\rm small\ circle} A \to {1\over
4\pi}\int_0^{2\pi}{R\over 2}d\tilde\varphi=R/4} where we have
computed the line integral in the $\rho\to 0$ limit.

Using the general method of  \EmparanBB, one can remove the dilaton from
\bigsolution\ to get a solution of the pure Einstein-Maxwell
theory.  In this manner we find
\eqn\biggersolution{\eqalign{
ds_{\rm EM}^2
&=\big(e^X+e^{-X}\rho^2/R^2\big)^2\big(-d\tau^2+e^{-4X}e^{8\gamma}(d\rho^2+dz^2)\big)
+\big(e^X+e^{-X}\rho^2/R^2\big)^{-2}\rho^2d\tilde\varphi^2\cr
A_{\rm EM}&=2A.}}
This solution describes a static linear array of extremal
Reissner-Nordstr\o m black holes with
alternating charges $\pm R/2$. Again the solution is free of singularities
because the opposing forces on
each black hole cancel without the need for  struts.

\subsec{Properties}

The geometry \bigsolution\ has the symmetry
\eqn\ttmsymmm{X\to-X+2\log(\rho/R),\qquad \gamma\to\gamma-X+\log(\rho/R),
}
times charge conjugation.  The transformation of $X$ and $\gamma$ can be
interpreted using
\ttmsym,\ \densityint\ as turning off the rods that are present, and
turning on rods where there were none.  This however is equivalent
to to shifting $z\to z+2k$.  Thus \bigsolution\ has the
symmetry $z\to z+2k$ times charge conjugation.  The regions
on the $z$-axis where there used to be rods are identical to the
regions on the $z$-axis where there had been no rods.

Next we write down the $\rho\to\infty$
asymptotic geometry and use it to compute the magnetic flux through
the $(\rho,\tilde\varphi)$-plane.  Since $X\sim\log(\rho/R)$, from \gis\
we get $\gamma\sim {1\over 4}\log\rho+C$, for $C$ a constant.  We find
\eqn\ourrholimit{\eqalign{ ds^2&\sim \big({2\rho\over R}\big)^{1/2}
\Big(-dt^2+\left({R\over\rho}\right)^{1/2}e^{2C}(d\rho^2+dz^2)+{\rho R\over 2}d\tilde\varphi^2\Big)\cr
A&\sim {R\over 4}d\tilde\varphi\cr
e^{-4\phi/\sqrt{3}}&\sim
{2\rho\over R}.}}
The gauge field in the $\rho\to\infty$ limit is a Wilson
line. From Stokes' theorem we can compute the flux through the
$(\rho,\tilde\varphi)$-plane at $z=0$, to be $R\pi/2$.  After dividing
by $4\pi$, this becomes
the s-charge $Q_s=R/8$ of the double Wick rotated solution. In the computation of
the flux through the $(\rho,\tilde\varphi)$-plane for $z=k$, one
encounters the Dirac string at $\rho=0$  yielding a flux of  $-R\pi/2$.

Naively the solution is described by two parameters, $k$ and $R$.  Actually,
the two solutions $(k,R)$ and $(\kappa k,R)$ are identical; sending $\rho\to\kappa\rho$
and $z\to\kappa z$ in the second solution and using $\exp(X(\kappa\rho,\kappa z;\kappa k))=
\kappa\,\exp(X(\rho,z;k))$, we see that they differ only by a coordinate transformation.
So we only have one parameter, $R$, which is proportional to the charge.

\subsec{Double Wick Rotation}

The double Wick rotation \wick, \wickk\ turns the solutions \bigsolution\
and \biggersolution\ into s-branes in the usual way.
For example, the dilatonic array \bigsolution\ becomes
\eqn\asd{\eqalign{
ds^2&=(e^X+e^{-X}\rho^2/R^2)^{1/2}\left(dy^2+e^{-X}e^{2\gamma}(d\rho^2-dt^2)+
{\rho^2\over e^X+e^{-X}\rho^2/R^2}d\tilde\varphi^2\right)\cr
A_\mu dx^\mu &={\rho^2 R\over 2\rho^2+2R^2e^{2X}}d\tilde\varphi\cr
e^{-4\phi/\sqrt{3}}&=e^X+e^{-X}\rho^2/R^2.
}}
As above, this is a non-singular, time-dependent solution
describing the creation and subsequent decay of an unstable
1-brane in four dimensions.

\newsec{Embedding into String Theory}

In this section we comment on the potential relevance of these
solutions for string theory.

An unstable D-brane in string theory has an open string tachyon on
its world volume \SenMG. A process in which this tachyon is
condensed in the far future, but nears the top of its potential
for a finite amount of time, describes the creation/decay of an
unstable brane. The result is a spacelike region of finite time
duration in which open strings can exist, or an s-brane \GutperleAI.
These objects are of interest because they are spacelike, and
therefore time-dependent, versions of the usual string theory
D-branes and hence may provide a useful tool for investigations of
time-dependent processes in string theory.

The starting point for a concrete description of s-branes is the
observation \SenNU\ that the open string tachyon profile \eqn\tis{
T (X^0) = \lambda \cosh (X^0/\sqrt{\alpha'}), } where $X^0$ is the
timelike worldsheet boson, leads to a time-dependent boundary CFT
describing the creation and subsequent decay of an unstable brane.
The corresponding boundary state sources a closed string
configuration, which describes the linearized far field behavior
of closed string fields in the presence of a rolling tachyon. For
the critical value of $\lambda=\half$ this boundary state is
related by Wick rotation to an array of sD-branes (i.e. D-branes
with a transverse time direction) located at imaginary times $t
={i (2n-1) \sqrt{\alpha'} } $, where $n$ is an integer
\refs{\GaiottoRM,\MaloneyCK,\SenNU,\LambertZR}). The reason for
this is simple: the tachyon profile \tis\ may be Wick rotated $X^0
\to i X^0_E$ to give the boundary deformation $\lambda \cos
(X^0_E/\sqrt{\alpha'})$ describing a periodic array of sD-branes
at the critical coupling. Interestingly, despite the fact that
there are no D-branes for real values of time when
$\lambda=\half$, the CFT boundary state still describes a highly
nontrivial closed string configuration. In superstring theories,
these boundary states describe an alternating array of D-branes
located at $t = {i (4n -1) \sqrt{\alpha'}/\sqrt{2}}$ and
anti-D-branes at $t=i{ (4n+1) \sqrt{\alpha'}/\sqrt{2}}$. This
boundary state contains direct information only about the
linearized closed string fields, corresponding to the
linearization of the full non-linear solutions described above.

A significant generalization of this construction was found in
\GaiottoRM.  Forgetting its origin as a critical rolling tachyon
on an unstable brane, the sD-brane array can simply be viewed as a
concise rule for constructing a classical closed string field
configuration. This rule can be consistently generalized by moving
the positions of the sD-branes subject to certain restrictions
\GaiottoRM.  In particular one can change the spacing or number of
branes.

The relation of the s-brane solutions discussed herein to string
theory constructions should now be clear. Consider an unstable
D7-brane in IIA string theory, with an open string tachyon of the
form \tis\ at the critical value, which is an alternating
${\rm sD}6-{\rm s\bar D}6$ array. Now, using the analysis of \GaiottoRM,
deform the solution by increasing the imaginary-time spacing
between the sD-branes to $k \gg \sqrt{\alpha'}$. Next increase the
number of initial unstable D-branes to $N$, yielding an array of
large $N$ clusters of ${\rm sD}6$ and ${\rm s\bar D}6$ branes.  In the limit of large $N$
with $g_sN$ fixed, this should have a dual gravity description as
the product of the 4D s-brane solution \bigsolution\ with
${\bf R}^6$.\foot{With a flat metric in the M-theory frame. Note that
D6-branes lift to KK monopoles in M-theory.}

Note that in motivating this correspondence we have only invoked
the standard and very general large $N$ open-closed string
duality. In particular we have not introduced any analog of the
near-horizon scaling limit \MaldacenaRE, which transforms general
open-closed string duality into a powerful and practical tool. It
would certainly be of great interest to find an analog of this
near-horizon scaling in the present time-dependent context.

Among the most interesting sD-brane arrays are those that arise
directly from tachyon profiles like \tis, without deforming the
spacing. In this case the sD-branes are separated by imaginary
distances of order the string scale -- in the closed string
description, this means that massive string states are becoming
important \LambertZR.\foot{To be more specific, the CFT
description of rolling tachyons describes Euclidean branes and
anti-branes separated by the imaginary distance $\sqrt{2\alpha'}$.
At this distance, the open string tachyon stretching between a
brane/anti-brane pair becomes massless, leading to the usual
infrared divergence in the partition function. In the closed
string language this is interpreted as an ultraviolet divergence
coming from very massive string states. } Hence the gravity
solutions described here are not trustworthy for such
configurations. Nevertheless a study of the limit of small spacing
could be interesting.

Finally we note that the constraints of \GaiottoRM\ allow a pair
of oppositely charged sD-branes at equal imaginary distances from
the real time axis.  Hence, following the preceding discussion,
one can find  an embedding of the black dihole s-brane solution
\sbd\ into string theory.

\centerline{\bf Acknowledgements}
This work was supported in part by DOE grants DE-FG02-91ER40654 and
DE-AC03-76SF0015.
We are grateful to
G. Gibbons, G. Horowitz, F. Leblond, E. Silverstein, and X. Yin for
useful conversations.

\listrefs

\end